# Thesis Report: Resource Utilization Provisioning in MapReduce

*Hamidreza Barati, Nasrin Jaberi*

*Department of IT and Computer Engineering*
*Payame Noor University*
*Nasrinjaberi60@yahoo.com*

**Abstract-** In this thesis report, we have a survey on state-of-the-art methods for modelling resource utilization of MapReduce applications regard to its configuration parameters. After implementation of one of the algorithms in literature, we tried to find that if CPU usage modelling of a MapReduce application can be used to predict CPU usage of another MapReduce application.

## 1. Introduction

MapReduce, introduced by Google in [1], has been known as a large-scale data processing technique indicating that both resource utilization and execution time are the most critical aspects of running a MapReduce application. As a result, provisioning resource usage and execution time of a MapReduce application becomes important before actual running of the application to assign proper time and resources. Also, MapReduce has been known as a parametric model which means severalparameters on cluster must be tuned before running an application. Generally, these parameters are assigned default values but researches in [2-3] show that the proper tuning of the parameters can make a MapReduce application run faster or with less resource utilization. Also, these researches indicate that the proper values of these parameters are application-dependent meaning that these values from one application to another application may change.

Among parameters influencing the performance of MapReduce cluster, in this study we will focus on the influences of four major parameters: size of input file, size of blocks, number of mappers and number of reducers.

Early works on modeling performance of MapReduce application was proposed in [4] for modeling the total execution time of Hadoop Hive queries which is based on using Kernel Canonical Correlation Analysis to obtain correlation between the performance feature vectors extracted from Map-Reduce job logs, and map time, reduce time, and total execution time. These features were acknowledged as critical characteristics for establishing any scheduling decisions. Another MapReduce modeling of computation utilizations was presented in [5-6]. After modeling each map and reduce phases independently by using dynamic linear programming, these modellings are combined to establish a global optimal strategy for MapReduce scheduling and resource allocation. In

In [2, 7], linear regression were used to model execution time and the total number of CPU tick clocks an application, respectively, regard to a few MapReduce parameters. After following the same modeling concept in this report, we are going to study if two applications may have the same model or not. In another word, (1) if prediction model of an application can be applied to predict the performance of another model or not and (2) if yes, which characteristics these applications should have?

## 2. The Resource Utilization Modelling in MapReduce

The primary idea, originally coming from [2, 7-9], is to model the computation cost (or CPU usage) of applications in Hadoop, pseudo-distributed mode. The general idea is to run an application (e.g.WordCount) for different values of 1) the number of Mappers, 2) the number of reducers, 3) The size of File System and 4) The size of input data. During running of the application for specific values of parameters, the real values of CPU usage are extracted from system (in Linux-Ubuntu). After 100 times running of this application for different values of these parameters (the values are chosen randomly in valid ranges), the relation between CPU usage and the four parameters is modelled by polynomial regression. The model is:

$$CPU\_Model = \alpha_0 + \alpha_1(Map) + \alpha_2(\text{Re}duce) + \alpha_3(FS\_Size) + \alpha_4(IN\_Size) + \alpha_5(Map)^2 + \alpha_6(\text{Re}duce)^2 + \alpha_7(FS\_Size)^2 + \alpha_8(IN\_Size)^2$$

Fig.1 shows the real CPU usage and the output of model for the experiments. Fig.2 is error between these two graphs (The error mean is 3.57%).

One question, which is the motivation behind this thesis, is that if the prediction model of an application is applicable for other applications? In which conditions two applications can be predicted with the same model? Primarily, two methods have been introduced in [10-11] to find the similarity between MapReduce applications by comparing the CPU utilization time series of these applications. The former one uses a mixture of Dynamic Time Warping and correlation analysis to find this similarity while the latter one uses statistical analysis to calculate minimum distance between applications' time series. These approaches conclude that if two applications have similar CPU patterns for several experiments with different values of parameters, it is more likely that the optimal values of configuration parameters for a particular application can be applied to optimally run another application. However, these methods do not study if prediction model can be shared between two similar applications or not. To show the problem more clear, we have examined another application (Exim Mainlog parsing) and calculated the difference between the actual CPU usage of this application and the prediction model of WordCount application. As can be seen from Fig.3 and 4, WordCount model can also predict Exaim Mainlog CPU usage with a good accuracy.

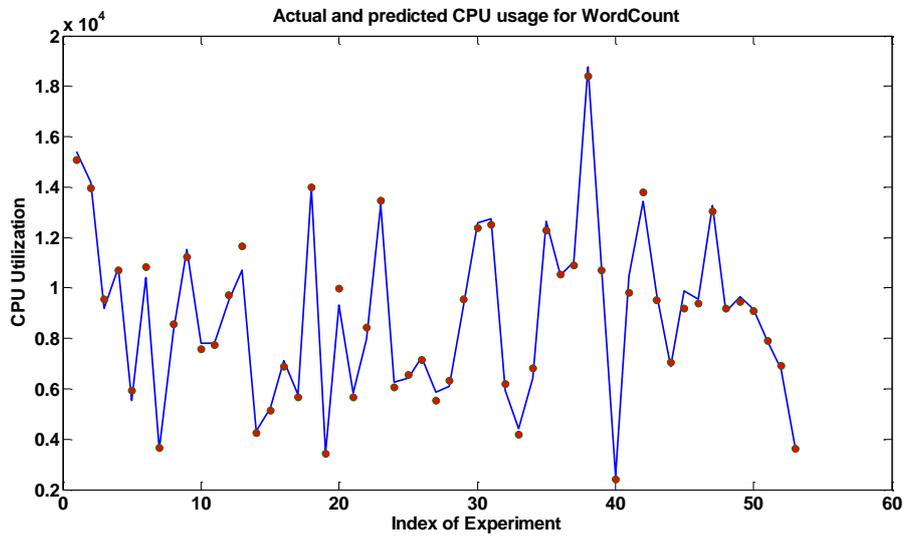

*Fig.1*

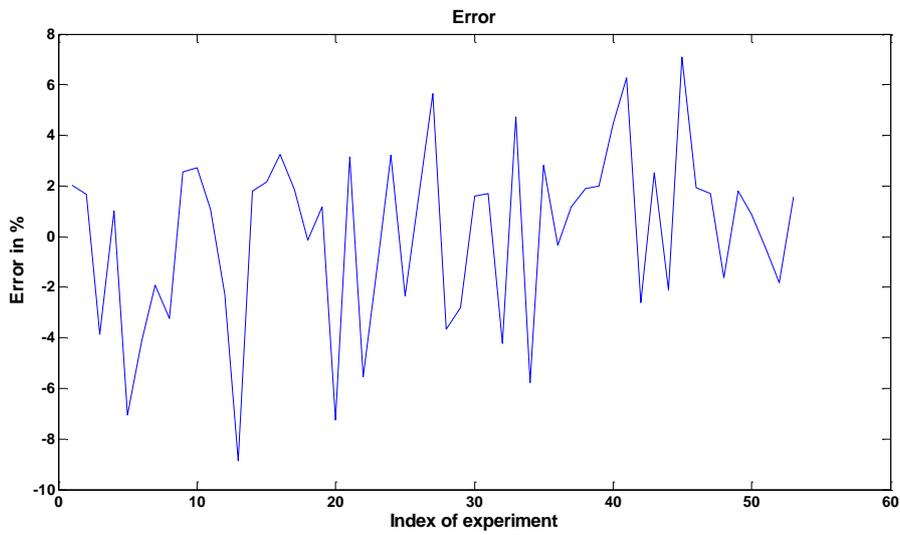

Fig .2

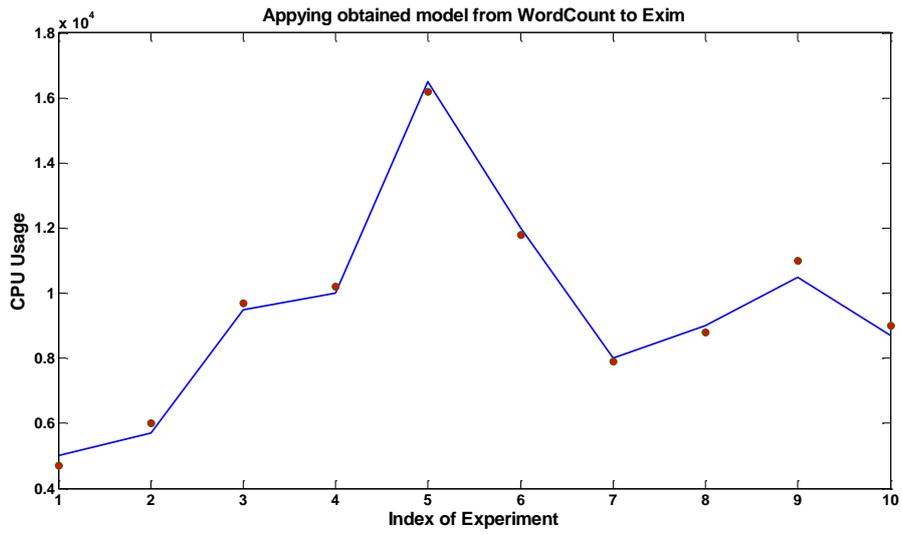

*Fig.3*

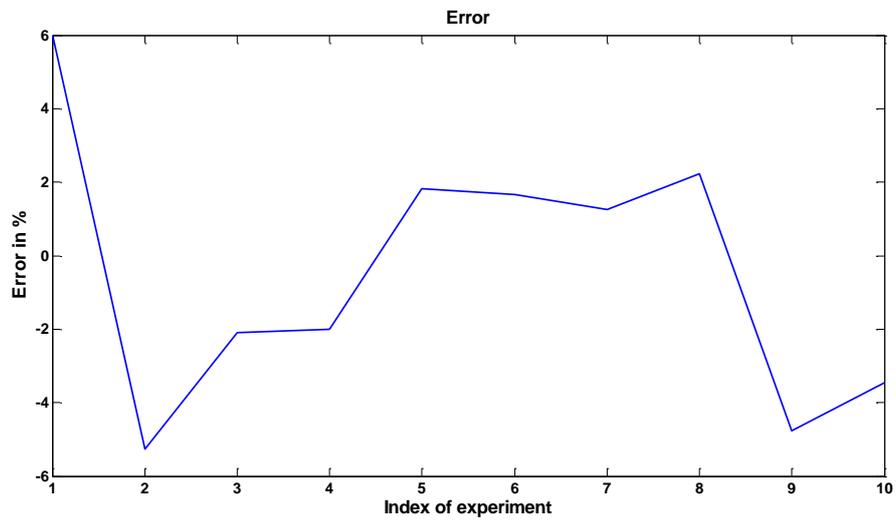

*Fig.4*